\documentclass[11pt]{article}
\usepackage{amsmath,amssymb}
\begin{document}
\title{{\bf On supremum of bounded quantum observable}\thanks{This project is supported by Natural Science Found of China (10771191 and
10471124).}}
\author {{Liu Weihua,\,\,  Wu Junde\thanks{E-mail: wjd@zju.edu.cn}}\\
{\small\it Department of Mathematics, Zhejiang University, Hangzhou
310027, P. R. China}}

\date{}
\maketitle

{\small\it \noindent Abstract. In this paper, we present a new
necessary and sufficient condition for which the supremum $A\vee B$
exists with respect to the logic order $\preceq$. Moreover, we give
out a new and much simpler representation of $A\vee B$ with respect
to $\preceq$, our results have nice physical meanings.}

\vskip0.2in


\vskip0.1in


\vskip0.1in

\section{Introduction}

There some basic notations: $H$ is a complex Hilbert space, $S(H)$
is the set of all bounded linear self-adjoint operators on $H$,
$S^{+}(H)$ is the set of all positive operators in $S(H)$, $P(H)$ is
the set of all orthogonal projection operators on $H$, ${\cal
B}(\mathbb{R})$ is the set of all Borel subsets of real number set
$\mathbb{R}$. Each element in $P(H)$ is said to be a quantum event
on $H$. Each element in $S(H)$ is said to be a bounded quantum
observable on $H$. For $A\in S(H)$, let $R(A)$ be the range of $A$,
$\overline {R(A)}$ be the closure of $R(A)$, $P_A$ be the orthogonal
projection on $\overline {R(A)}$, $P^{A}$ be the spectral measure of
$A$, null$(A)$ be the null space of $A$, and $N_A$ be the orthogonal
projection on null$(A)$.

Let $A, B\in S(H)$. If for each $x\in H$, $[Ax, x]\leq [Bx, x]$,
then we say that $A\leq B$. Equivalently, there exists a $C\in
S^{+}(H)$ such that $A+C=B$. $\leq $ is a partial order on $S(H)$.
The physical meaning of $A\leq B$ is that the expectation of $A$ is
not greater than the expectation of $B$ for each state of the
system. So the order $\leq $ is said to be a numerical order of
$S(H)$. But $(S(H), \leq)$ is not a lattice. Nevertheless, as a well
known theorem due to Kadison, $(S(\mathbb{H}), \leq)$ is an
anti-lattice, that is, for any two elements $A$ and $B$ in
$S(\mathbb{H})$, the infimum $A\wedge B$ of $A$ and $B$ exists with
respect to $\leq$ iff $A$ and $B$ are comparable with respect to
$\leq$ ([1]).

In 2006, Gudder introduced a new order $\preceq$ on $S(H)$: if there
exists a $C\in S(H)$ such that $AC=0$ and $A+C=B$, then we say that
$A\preceq B$ ([2]).

Equivalently, $A\preceq B$ iff for each $\Delta\in{\cal
B}(\mathbb{R})$ with $0\notin \Delta$, $P^{A}(\Delta)\leq
P^{B}(\Delta)$ ([2]). The physical meaning of $A\preceq B$ is that
for each $\Delta\in{\cal B}(\mathbb{R})$ with $0\notin \Delta$, the
quantum event $P^{A}(\Delta)$ implies the quantum event
$P^{B}(\Delta)$. Thus, the order $\preceq $ is said to be a logic
order of $S(H)$ ([2]). In [2], it is proved that $(S(H),\preceq)$ is
not a lattice since the supremum of arbitrary $A$ and $B$ may not
exist in general. In [3], it is proved that the infimum $A\wedge B$
of $A$ and $B$ with respect to $\preceq$ always exists. In [4, 5],
the representation theorems of the infimum $A\wedge B$ of $A$ and
$B$ with respect to $\preceq$ were obtained. In more recent, Xu and
Du and Fang in [6] discussed the existence of the supremum $A\vee B$
of $A$ and $B$ with respect to $\preceq$ by the technique of
operator block. Moreover, they gave out a sufficient and necessary
conditions for the existence of $A\vee B$ with respect to $\preceq$.
Nevertheless, their conditions are difficult to be checked since the
conditions depend on an operator $W$, but $W$ is not easy to get.
Moreover, their proof is so much algebraic that we can not
understand its physical meaning.

In this paper, we present a new necessary and sufficient condition
for which $A\vee B$ exists with respect to $\preceq$ in a totally
different form. furthermore, we give out a new and much simpler
representation of $A\vee B$ with respect to $\preceq$, our results
have nice physical meanings.

{\bf Lemma 1.1} [2]. Let $A,B\in S(H)$. If $A\preceq B$, then
$A=BP_A$.

{\bf Lemma 1.2} [2]. If $P, Q\in P(H)$, then $P\leq Q$ iff $P\preceq
Q$, and $P$ and $Q$ have the same infimum $P\wedge Q$ and the
supremum $P\vee Q$ with respect to the orders $\leq $ and $\preceq
$, we denote them by $P\wedge Q$ and $P\vee Q$, respectively.

{\bf Lemma 1.3} [7]. Let $A, B\in S(H)$. Then $P^A(\{0\})=N(A)$,
$P_A=P^A(R\backslash\{0\})$, $P_A+N(A)=I$, $P_A\vee P_B=I-N(A)\wedge
N(B)$.

\section{Some elementary lammas}

Let $A,B\in S(H)$ and they have the following forms:
$$A=\int\limits_{-M}^{M} \lambda dA_{\lambda}$$ and
 $$B=\int\limits_{-M}^{M} \lambda dB_{\lambda},$$ where $\{A_{\lambda}\}_{\lambda\in\mathbb{R}}$
and $\{B_{\lambda}\}_{\lambda\in\mathbb{R}}$ be the identity
resolutions of $A$ and $B$ ([7]), respectively, and $M=\max(\|A\|,
\|B\|)$.

If $A$ has an upper bound $F$ in $S(H)$ with respect to $\preceq$,
then it follows from Lemma 1.1 that $A=FP_A$. Note that $A\in S(H)$,
so $FP_A=P_AF$ and thus $AF=FA$. Let $F$ have the following form:
$$F=\int\limits_{-G}^{G} \lambda dF_{\lambda},$$
where $\{F_{\lambda}\}_{\lambda\in\mathbb{R}}$ is the identity
resolution of $F$ and $G=\max(\|F\|,M)$. Then we have
$$A=FP_A=(\int\limits_{-G}^{G} \lambda dF_{\lambda})P_A
=\int\limits_{-G}^{G} \lambda d(F_{\lambda}P_A).$$

{\bf Lemma 2.1.} Let $A\in S(H)$ and $F\in S(H)$ be an upper bound
of $A$ with respect to $\preceq$. Then for each $\Delta\in {\cal
B}(\mathbb{R})$, we have
$$P^A(\Delta)=\left\{
\begin{array}{ccc}
    P^F(\Delta)P_A, &\ \ & 0\not\in\Delta  \\
    N(A), &\ \ &\Delta=\{0\}\\
    P^F(\Delta\backslash\{0\})P_A+N(A). &\ \ & 0\in\Delta \\
\end{array}\right.
$$

{\bf Proof.} We just need to check $P^A(\Delta)=P^F(\Delta)P_A$ when
$0\not\in\Delta$, the rest is trivial. Note that if we restrict on
the subspace $P_A(H)=\overline {R(A)}$, since $AF=FA$, then
$\{F_{\lambda}P_A\}_{\lambda\in\mathbb{R}}$ is the identity
resolution of $F|_{P_A(H)}$ ([7]). Let $f$ be the characteristic
function of $\Delta$. Then the following equality proves the
conclusion:
$$P^A(\Delta)=f(A)=f(FP_A)=\int\limits_{-G}^{G} f(\lambda) d(F_{\lambda}P_A)
=\int\limits_{\lambda\in \Delta}d(F_{\lambda}P_A)=P^F(\Delta)P_A.$$

It follows from Lemma 2.1 immediately:

{\bf Lemma 2.2.} Let $A, B\in S(H)$ and $F\in S(H)$ be an upper
bound of $A$ and $B$ with respect to $\preceq$. Then for any two
Borel subsets $\Delta_1$ and $\Delta_2$ of $\mathbb{R}$, if
$\Delta_1\cap\Delta_2=\emptyset$, $0\notin\Delta_1$,
$0\notin\Delta_2$, we have
$$P^A(\Delta_1)P^B({\Delta_2})=P^F(\Delta_1)P_AP^F(\Delta_2)P_B=
P_AP^F(\Delta_1)P^F(\Delta_2)P_B=\theta.$$

{\bf Lemma 2.3.} Let $A, B\in S(H)$ and have the following property:
For each pair $\Delta_1, \Delta_2\in {\cal B}(\mathbb{R})$, whenever
$\Delta_1\cap\Delta_2=\emptyset$ and $0\not\in\Delta_1$,
$0\not\in\Delta_2$, we have $P^A(\Delta_1)P^B({\Delta_2})=\theta$,
then the following mapping $E: {\cal B}(\mathbb{R})\rightarrow P(H)$
defines a spectral measure:
$$
E(\Delta)=\left\{ \begin{array}{ccc}
    P^A(\Delta)\vee P^B(\Delta), &\ \ & 0\not\in\Delta  \\
    N(A)\wedge N(B)=I-P_A\vee P_B, &\ \ &\Delta=\{0\}\\
     P^A(\Delta\backslash\{0\})\vee P^B(\Delta\backslash\{0\})+N(A)\wedge N(B). &\ \ & 0\in\Delta \\
\end{array}\right.$$

{\bf Proof.} First, we show that for each $\Delta\in {\cal
B}(\mathbb{R})$, $E(\Delta)\in P(H)$. It is sufficient to check the
case of $0\in \Delta$. Since $P^A(\Delta\backslash\{0\})\vee
P^B(\Delta\backslash\{0\})\leq P^A(R\backslash\{0\})\vee
P^B(R\backslash\{0\})=P_A\vee P_B$, so it follows from Lemma 1.3
that $P^A(\Delta\backslash\{0\})\vee
P^B(\Delta\backslash\{0\})+N(A)\wedge N(B)\in P(H)$ and the
conclusion is hold.

Second, we have $$E(\emptyset)=P^A(\emptyset)\vee
P^B(\emptyset)=\theta\vee \theta=\theta,$$
$$E(R)=P^A(R\backslash\{0\})\vee P^B(R\backslash\{0\})+N(A)\wedge
N(B)$$$$=P_A\vee P_B+N(A)\wedge N(B)=I.$$

Third, if $\Delta_1\cap\Delta_2=\emptyset$, there are two cases:

(i). $0$ doesn't belong to any one of $\Delta_1$ and $\Delta_2$. It
follows from the definition of $E$ that $
E(\Delta_1)E(\Delta_2)=(P^A(\Delta_1)\vee
P^B(\Delta_1))(P^A(\Delta_2)\vee P^B(\Delta_2)).$ Note that
$P^B(\Delta_1)P^A(\Delta_2)=\theta$ by the conditions of the lemma
and $P^B(\Delta_1)P^B(\Delta_2)=\theta$, we have
$P^B(\Delta_1)(P^A(\Delta_2)\vee P^B(\Delta_2))=\theta$, similarly,
we have also $P^A(\Delta_1)(P^A(\Delta_2)\vee
P^B(\Delta_2))=\theta$, thus, $$E(\Delta_1)E(\Delta_2)=\theta.$$

\vskip0.1in

Furthermore, we have

\vskip0.1in

$\begin{array}{rcl}
E(\Delta_1\cup\Delta_2)&=&P^A(\Delta_1\cup\Delta_2)\vee
P^B(\Delta_1\cup\Delta_2)\\
&=&P^A(\Delta_1)\vee P^A(\Delta_2)\vee P^B(\Delta_1)\vee
P^B(\Delta_2)\\&=&(P^A(\Delta_1)\vee P^B(\Delta_1))\vee
(P^A(\Delta_2)\vee P^B(\Delta_2))\\&=&(P^A(\Delta_1)\vee
P^B(\Delta_1))+ (P^A(\Delta_2)\vee
P^B(\Delta_2))\\&=&E(\Delta_1)+E(\Delta_2).\end{array}$

That is, in this case, we proved that
$$E(\Delta_1)E(\Delta_2)=\theta,$$ $$E(\Delta_1\cup\Delta_2)=E(\Delta_1)+E(\Delta_2).$$

(ii). $0$ belongs to one of $\Delta_1$ and $\Delta_2$. Without of
 losing generality, we suppose that $0\in \Delta_1$, since $\Delta_1\cap\Delta_2=\emptyset$, so $0\notin \Delta_2$, thus
 we have

 \vskip0.1in

 $\begin{array}{rcl}
    E(\Delta_1)E(\Delta_2)&=&(P^A(\Delta_1\backslash\{0\})\vee
P^B(\Delta_1\backslash\{0\})+N(B)\wedge N(A))(P^A(\Delta_2)\vee P^B(\Delta_2))\\
   &=&(P^A(\Delta_1\backslash\{0\})\vee P^B(\Delta_1\backslash\{0\}))(P^A(\Delta_2)\vee P^B(\Delta_2))=\theta,\\
\end{array}$\\

\vskip0.1in

$\begin{array}{rcl}
    E(\Delta_1\cup\Delta_2)&=&P^A(\Delta_1\backslash\{0\}\cup\Delta_2)\vee
P^B(\Delta_1\backslash\{0\}\cup\Delta_2)+(N(B)\wedge
N(A))\\
&=&(P^A(\Delta_1\backslash\{0\})\vee
P^B(\Delta_1\backslash\{0\})+(N(B)\wedge N(A)))+(P^A(\Delta_2)\vee P^B(\Delta_2))\\
   &=&(P^A(\Delta_1\backslash\{0\})\vee
P^B(\Delta_1\backslash\{0\})+(N(A)\wedge N(B)))+(P^A(\Delta_2)\vee P^B(\Delta_2))\\
&=&E(\Delta_1)+E(\Delta_2).\\
\end{array}$

\vskip0.1in

Thus, it follows from (i) and (ii) that whenever
$\Delta_1\cap\Delta_2=\emptyset$, we have
$$E(\Delta_1)E(\Delta_2)=\theta,$$
$$E(\Delta_1\cup\Delta_2)=E(\Delta_1)+E(\Delta_2).$$

\vskip0.1in

Final, if $\{\Delta_n\}_{n=1}^{\infty}$ is a sequence of pairwise
disjoint Borel sets in ${\cal B}(\mathbb{R})$, then it is easy to
prove that
$$E(\bigcup\limits_{n=1}^{\infty}\Delta_n)=\sum\limits_{n=1}^{\infty}E(\Delta_n).$$

Thus, the lemma is proved.

\section{Main results and proofs}

{\bf Theorem 3.1.}  Let $A, B\in S(H)$ and have the following
property: For each pair $\Delta_1, \Delta_2\in {\cal
B}(\mathbb{R})$, whenever $\Delta_1\cap\Delta_2=\emptyset$ and
$0\not\in\Delta_1$, $0\not\in\Delta_2$, we have
$P^A(\Delta_1)P^B({\Delta_2})=\theta$. Then the supremum $A\vee B$
of $A$ and $B$ exists with respect to the logic order $\preceq$.

{\bf Proof.} By Lemma 2.3, $E(\cdot)$ is a spectral measure and so
it can generate a bounded quantum observable $K$ and $K$ can be
represented by $K=\int\limits_{-M}^{M} \lambda dE_{\lambda}$, where
$\{E_{\lambda}\}=E(-\infty, \lambda]$, $\lambda\in\mathbb{R}$ and
$M=\max(\|A\|, \|B\|)$. Moreover, for each $\Delta\in {\cal
B}(\mathbb{R})$, $P^K(\Delta)=E(\Delta)$ ([7]). We confirm that $K$
is the supremum $A\vee B$ of $A$ and $B$ with respect to $\preceq$.
In fact, for each $\Delta\in {\cal B}(\mathbb{R})$ with
$0\notin\Delta$, by the definition of $E$ we knew that
$P^K(\Delta)=E(\Delta)=P^A(\Delta)\vee P^B(\Delta)\geq P^A(\Delta)$,
 $P^K(\Delta)=E(\Delta)=P^A(\Delta)\vee P^B(\Delta)\geq P^B(\Delta)$. So it following from the equivalent properties of $\preceq$ that $A\preceq K$, $B\preceq K$ ([2]). If $K'$ is another upper bound
of $A$ and $B$ with respect to $\preceq$, then for each
$\Delta\in{\cal B}(\mathbb{R})$ with $0\notin \Delta$, we have
$P^A(\Delta)\leq P^{K'}(\Delta)$, $P^B(\Delta)\leq P^{K'}(\Delta)$
([2]), so $P^A(\Delta)\vee P^B(\Delta)=E(\Delta)=P^K(\Delta)\leq P^
{K'}(\Delta)$, thus we have $K\preceq K'$ and $K$ is the supremum of
$A$ and $B$ with respect to $\preceq$ is proved.

\vskip 0.1 in

It follows from Lemma 2.2 and theorem 3.1 that we have the following
theorem immediately:

{\bf Theorem 3.2.}  Let $A, B\in S(H)$. Then the supremum $A\vee B$
of $A$ and $B$ exists with respect to the logic order $\preceq$ iff
for each pair $\Delta_1, \Delta_2\in {\cal B}(\mathbb{R})$, whenever
$\Delta_1\cap\Delta_2=\emptyset$ and $0\not\in\Delta_1$,
$0\not\in\Delta_2$, we have $P^A(\Delta_1)P^B({\Delta_2})=\theta$.
Moreover, in this case, we have the following nice representation:
$$A\vee B=\int\limits_{-M}^{M} \lambda
dE_{\lambda},$$ where $\{E_{\lambda}\}=E(-\infty, \lambda]$,
$\lambda\in\mathbb{R}$ and $M=\max(\|A\|, \|B\|)$.

\vskip 0.1 in

{\bf Remark 3.3.} Let $A, B\in S(H)$. Note that for each $\Delta\in
{\cal B}(\mathbb{R})$, $P^A(\Delta)$ is interpreted as the quantum
event that the quantum observable $A$ has a value in $\Delta$ ([2]),
and the conditions: $\Delta_1\cap\Delta_2=\emptyset$,
$0\not\in\Delta_1$, $0\not\in\Delta_2$ must have
$P^A(\Delta_1)P^B({\Delta_2})=\theta$ told us that the quantum
events $P^A(\Delta_1)$ and $P^B(\Delta_2)$ can not happened at the
same time, so, the physical meanings of the supremum $A\vee B$
exists with respect to $\preceq$ iff for each pair $\Delta_1,
\Delta_2\in {\cal B}(\mathbb{R})$, whenever
$\Delta_1\cap\Delta_2=\emptyset$ and $0\not\in\Delta_1$,
$0\not\in\Delta_2$, the quantum observable $A$ takes value in
$\Delta_1$ and the quantum observable $B$ takes value in $\Delta_2$
can not happen at the same time.

\vskip0.2in

\centerline{\bf References}

\vskip0.2in

\noindent [1]. Kadison, R. Order properties of bounded self-adjoint
operators.  \emph{Proc. Amer. Math. Soc}. {\bf 34}: 505-510, (1951)

\noindent [2]. Gudder S. An Order for quantum observables.
\emph{Math Slovaca}. {\bf 56}: 573-589,  (2006)

\noindent [3]. Pulmannova S, Vincekova E. Remarks on the order for
quantum observables. \emph{Math Slovaca}. {\bf 57}: 589-600, (2007)

\noindent [4]. Liu Weihua, Wu Junde. A representation theorem of
infimum of bounded quantum observables. \emph{J Math Physi}. {\bf
49}: 073521-073525, (2008)

\noindent [5]. Du Hongke, Dou Yanni. A spectral representation of
infimum of self-adjoint operators in the logic order. Acta Math.
Sinica. To appear

\noindent [6]. Xu Xiaoming, Du Hongke, Fang Xiaochun. An explicit
expression of supremum of bounded quantum observables. \emph{J Math
Physi}. {\bf 50}: 033502-033509, (2009)

\noindent [7]. Kadison. R. V., Ringrose J. R. {\it Fundamentals of
the Theory of Operator Algebra.} Springer-Verlag, New York, (1983)

\end{document}